\documentclass[12pt,preprint]{aastex}
\usepackage{epsfig}

\shortauthors{Cui \& Konopelko}
\shorttitle{X-ray Counterpart of HESS~J1804-216}

\def\be{\begin{equation}}
\def\ee{\end{equation}}

\def\be{\begin{equation}}
\def\ee{\end{equation}}
\catcode`\@=11 
\def\@versim#1#2{\vcenter{\offinterlineskip
\ialign{$\m@th#1\hfil##\hfil$\crcr#2\crcr\sim\crcr } }} 
\def\lsim{\mathrel{\mathpalette\@versim<}}
\def\gsim{\mathrel{\mathpalette\@versim>}}

\begin{document}

\title{Chandra View of the Unidentified TeV Gamma-ray Source HESS~J1804-216}

\author{Wei Cui and Alexander Konopelko}
\affil{Department of Physics, Purdue University, West Lafayette, IN 47907}

\begin{abstract}

We present high-resolution X-ray images taken with the {\em Chandra X-ray
Observatory} of the field that contains the unidentified TeV gamma-ray source 
HESS~J1804-216. A total of eleven discrete sources were detected with {\it
a posteriori} significance of $> 5\sigma$ over the entire field of view. 
Among them, only one, designated as CXOU~J180351.4-213707, is significantly 
extended. The source is about 40\arcsec\ away from the radio pulsar 
PSR~J1803-2137, which was the target of the {\em Chandra} observation
but was not detected in X-rays. A natural question is whether the two 
sources are physically related. While it is conceivable that 
CXOU~J180351.4-213707 could be associated with a previously unknown supernova 
remnant (SNR), in which the pulsar was born, it seems equally plausible that 
it might be a pulsar wind nebula (PWN) that is powered by a different
pulsar whose emission is beamed away from us. In either case, we argue that 
CXOU~J180351.4-213707 is likely the X-ray counterpart of HESS~J1804-216, 
based on the fact that the Galactic TeV gamma-ray sources are predominantly 
SNRs or PWNe. The X-ray spectrum of the source can be fitted well with a 
power law, although the model is not well constrained due to large statistical
uncertainties. The spectrum seems to be very hard, with the best-fit photon 
index $\sim 1.2$. Under the assumption that CXOU~J180351.4-213707 is 
the X-ray counterpart of HESS~J1804-216, we attempted to model the X-ray and 
TeV emission as synchrotron and inverse Compton scattered radiation from 
relativistic electrons. We briefly discuss the results.

\end{abstract}

\keywords{acceleration of particles ---  gamma rays: theory --- pulsars: 
individual (PSR J1803-2137) --- radiation mechanisms: non-thermal 
--- supernova remnants ---  X-rays: general}

\section{Introduction}

One of the most exciting recent advances in high energy astrophysics is the 
detection of various classes of sources at TeV energies with ground-based 
gamma ray facilities (for recent reviews see Weekes 2006 and Cui 2006). The
established TeV gamma-ray emitters now include blazars, radio galaxies, 
shell-type supernova remnants (SNRs), pulsar wind nebulae (PWNe), 
microquasars, 
and Be binaries, all of which have traditionally been standard targets for 
X-ray/soft gamma-ray observations. Arguably the most significant discovery 
is, however, the presence of a population of unidentified TeV gamma-ray 
sources. These sources are probably Galactic in origin, given their 
concentration around the Galactic plane (though one must take into account 
a strong observational bias towards the Galactic plane). Some of them have 
plausible counterparts at longer wavelengths (Aharonian et al. 2005a, 2006a), 
based mostly on positional coincidence, but others have none at all 
(Mukherjee et al. 2003; Butt et al. 2003; Aharonian et al. 2005a, 2006a).

A number of proposals have been made on the nature of the unidentified TeV 
gamma-ray sources. The sources might be associated with old shell-type SNRs 
(Yamazaki et al. 2006). Such systems could have a very high ratio of TeV to 
X-ray (or radio) fluxes, which makes them difficult to detect at low energies.
For instance, HESS J1813-178 was initially ``dark'' (Aharonian et al. 2005a) 
but was subsequently identified with a shell-type SNR (G12.8-0.0; Brogan et 
al. 2005). On the other hand, HESS J1825-137 is now positively associated 
with a PWN (G18.0-0.7; Aharonian et al. 2006b). Therefore, both SNRs and PWNe 
are viable candidates. It is worth noting that the plausible counterparts of 
unidentified TeV gamma-ray sources are mostly shell-type SNRs or PWNe 
(Aharonian et al. 2005a, 2006a). Other possibilities also exist. For instance,
HESS J1303-631 is postulated as the remnant of a gamma-ray burst that occurred
some tens of thousands of years ago in our galaxy (Atoyan et al. 2006). 
TeV~J2032+4130 might be associated with shocks produced by colliding winds of 
massive stars in the dense Cyg OB2 association (Butt et al. 2003).

In this Letter, we report the detection of extended X-ray emission near the
radio pulsar PSR~J1803-2137. The pulsar has been suggested as a possible
counterpart of HESS~J1804-216, one of the brightest unidentified TeV gamma-ray
sources (Aharonian et al. 2005a,2006a). While our results cannot definitively 
establish a physical connection between the extended emission and the pulsar, 
they have provided evidence to suggest that the former may be the X-ray 
counterpart of HESS~J1804-216.

\section{Data Analysis and Results}

The data for this work were derived from an archival {\it Chandra} observation
of PSR J1803-2137 (ObsID \#5590). The data were taken with the ACIS detector,
with CCDs I2--3 and S0--3 being read out. The aim point is on the S3 chip 
(with the default Y offset $\Delta Y$=$-20$\arcsec). The total on-source time 
is about 30 ks. The data were reduced with the standard {\it CIAO} analysis 
package (version 3.3), along with CALDB 3.2.0. We followed the {\it CIAO} 
Science Threads\footnote{see http://asc.harvard.edu/ciao/threads/index.html} 
in preparing the data (including bad pixel removal and data filtering) and 
constructing images and spectra. Little data were filtered out in this case, 
so the effective exposure time is also about 30 ks. We found that the Level 2 
event file (with a processing version of 7.6.7) from the archive had already 
incorporated the most updated calibrations, so we used it as a starting point 
for subsequent imaging and spectral analysis.

\subsection{Imaging Analysis}

We made an X-ray image of the field in the 0.5--10 keV band and carried out a 
search for discrete sources. The {\it CIAO} tool {\it celldetect} was used. 
It is based on the sliding-cell algorithm, but the detection cell is allowed
to vary in size to match the local point spread function (PSF). The key 
parameters are the signal-to-noise (S/N) threshold and the size of the 
detection cell, 
which are, by default, set at 3 and 80\% of the encircled energy of the PSF, 
respectively. We also left all other parameters at the default values. The
default settings have been shown to be effective against spurious 
detections.\footnote{see http://asc.harvard.edu/ciao/download/doc/detect\_manual/cell\_false.html.} We assessed the statistical significance of each 
detection with Eq. (17) in Li \& Ma (1983), based on the output of 
{\it celldetect} (namely, the sizes of the source and background regions 
and the total counts in the two regions). Table~1 shows all detections with 
significance $\gtrsim 5\sigma$. Seven of the sources are on the S3 chip, 
two on the S2 chip, two on the S1 chip, and one on the I3 chip. 

A few remarks on the results are necessary. First of all, only statistical 
uncertainties are shown for the best-fit positions of the sources. It is 
known that the systematic uncertainty on the absolute position of a source 
can be much larger, about 0.6\arcsec\ in radius of the 90\% error circle and 
0.8\arcsec\ of the 99\% error circle on average (but larger for sources more 
than 3\arcmin\ from the aim point)\footnote{See http://asc.harvard.edu/cal/ASPECT/celmon/}. Secondly, for extended sources, {\it celldetect} only determines 
the coordinates of the centroid. Thirdly, the count rates shown are also from 
{\it celldetect}, except for CXOU J180351.4-213707, which {\it celldetect} 
failed to adequately characterize, due to its extended nature. To accurately 
extract source counts from CXOU J180351.4-213707, we used a circular source 
region that centered on it with a radius of 10 pixels (or about 5\arcsec) 
and a concentric background annulus with an inner radius of 10 pixels and 
an outer radius of 25 pixels (chosen to avoid a possible source that is 
below our detection threshold; see Fig. 1). The true count rate of the 
source is shown in Table~1.

Among the sources that we detected, CXOU J180351.4-213707 is the only one 
that is significantly extended. Figure~1 shows an expanded view of the source,
along with a broad view of PSR~J1803-2137 and its surroundings. The image has 
been smoothed (in {\it ds9} ) with a Gaussian kernel that is 3 pixels in 
radius. CXOU J180351.4-213707 is seen to extend roughly 7\arcsec\ along both 
right accession and declination. It is about 40\arcsec\ away from 
PSR~J1803-2137. No X-ray 
emission from the latter is detected. To be more quantitative, we extracted 
counts from a circular source region that is 5 pixels in radius and centered 
at the position of the pulsar and from a concentric background annulus with 
an inner radius of 10 pixels and outer radius of 40 pixels. The resulting 
net count rate is $(-5\pm 9)\times 10^{-5}$ $cts$ $s^{-1}$.

We searched the {\it SIMBAD} and {\it NED} databases for plausible 
counterparts of the detected sources. Within a 30\arcsec\ search radius, 
we found only one candidate, 1WGA J1803.6-2140, which is about 12\arcsec\ 
away from CXOU J180341.5-214034. In addition, CXOU J180432.4-214009 appears 
to be the same source as Suzaku J1804-2140 (Bamba et al. 2006), which was 
suggested as a plausible X-ray counterpart of HESS J1804-216, although the
source is not obviously extended in our case. We should note that this 
source falls on the I3 chip, which is quite far from the aim point. No 
X-ray emission was detected at the position of Suzaku J1804-2142 (Bamba et 
al. 2006), implying that it is either a transient or highly variable 
source. 

\subsection{Spectral Analysis}

We used the {\it CIAO} script {\it specextract} to construct the X-ray
spectrum of CXOU J180351.4-213707. Here, we adopted the same source 
region as before but used a much larger background annulus, the outer
radius of which is 170 pixels. We excluded a circular region (of 9-pixel
radius) that is centered on the possible source to the northwest of 
CXOU J180351.4-213707, just to be on the cautious side. The script 
produced both the overall and background spectra, as well as the 
corresponding {\it rmf} and {\it arf} files that are needed for subsequent 
spectral modeling.

For spectral analysis, we excluded data points below 0.3 keV and above 10 
keV and then rebinned the raw spectrum so that there are at least 15 counts 
in each bin. We carried out spectral modeling in {\it XSPEC 11.3.2} (Arnaud 
1996). The background-subtracted spectrum is shown in Figure~2. It can be 
fitted well with a simple absorbed power-law model ({\it wabs*powerlaw} in 
{\it XSPEC}). The best-fit model and residuals are also shown in Fig.~2. The 
reduced $\chi^2$ of the fit is 0.57 for 8 degrees of freedom. The derived 
parameters are: hydrogen column density $N_H=8^{+6}_{-3}\times 10^{21}$ 
$cm^{-2}$, photon index $\Gamma=1.2^{+0.5}_{-0.4}$, and normalization 
$K=1.1^{+1.2}_{-0.5}\times 10^{-5}$ $ph$ $cm^{-2}$ $s^{-1}$ $keV^{-1}$ 
at 1 keV. The errors shown represent 90\% confidence limits. Although the 
model is not well constrained, due to large statistical uncertainties, 
CXOU J180351.4-213707 seems to be a very hard X-ray source. The 
spatially-averaged 
(absorbed) flux of the source is $1.0^{+29.3}_{-0.9}\times 10^{-13}$ $erg$ 
$cm^{-2}$ $s^{-1}$ (in the 0.3--10 keV band).

\section{Discussion}

The {\it Chandra} observation has revealed the presence of eleven discrete
X-ray sources in the general vicinity of HESS J1804-216. Figure~3 shows 
the positions of these sources in galactic coordinates, as overlaid over
the TeV gamma-ray image of the field (Aharonian et al. 2006a). While it
is difficult to be certain as to which is the counterpart of HESS~J1804-216, 
we argue that CXOU~J180351.4-213707 is most probable. It is the only 
significantly extended source detected. It is conceivable that the source
might be a previously unknown SNR that is associated with PSR~J1803-2137. If 
so, the spatial offset between the two could be attributed to the proper 
motion of the pulsar. At the distance of the pulsar ($\sim 4$ kpc; Taylor 
\& Cordes 1993), the separation between the two is only about 0.78 pc. 
Assuming the pulsar is born at the ``center'' of CXOU J180351.4-213707, it 
would only require a transverse speed of $\lesssim 50$ km/s for it to reach 
the current position in $\sim$16,000 yrs, which is its characteristic 
spin-down age ($\equiv \frac{P}{2\dot{P}}$; Clifton \& Lyne 1986). This 
would be easily achievable. 
However, the measured size of CXOU~J180351.4-213707 is only about 0.14 pc 
(for an angular size of $\sim$7\arcsec), which seems to be much too small 
for an SNR that is over $10^4$ years old. 

On the other hand, {\em Chandra} might have revealed only the brightest part 
of the hypothesized SNR and its true size might be much larger. In fact, 
CXOU~J180351.4-213707 might only represent a bright spot of X-ray emission 
of such an SNR. This scenario can be tested with a much deeper exposure of 
the field with {\em Chandra} or {\em XMM-Newton} in the future. It should be 
noted that PSR J1803-2137 was initially thought to be associated with the 
SNR G8.7-0.1 (Kassim \& Weiler 1990; see also Fig.~3), based on similar 
estimated distances and ages 
between the two. It was, however, subsequently realized that an unusually 
large transverse velocity ($\sim 1700$ $km$ $s^{-1}$) would be required for 
the pulsar to reach the current position if it was born at the center of 
G8.7-0.1. This, along with a smaller revised distance to the pulsar and the 
lack of evidence for such a large transverse velocity, led to a 
dis-association between the two systems (Frail et al. 1994; see, however, 
Finley \& \"{O}gelman 1994). There would be no similar issues with the 
scenario that we are postulating here, because the pulsar might be even 
closer to the dynamical center of the hypothesized SNR than to 
CXOU~J180351.4-213707 (about 0.78 pc, at the distance of PSR~J1803-2137).

Alternatively, CXOU~J180351.4-213707 might have no physical connection with 
PSR~J1803-2137. Instead, it could be a PWN that is powered by a different
pulsar whose emission is beamed away from us. Again, the true size of such
a PWN could be much larger than what has been measured. For instance,
CXOU~J180351.4-213707 might be the bright compact core of the PWN. Such a 
core is known to exist in several PWNe. As an example, a torus-like compact 
X-ray emitting region was seen around the Vela pulsar with {\it Chandra} 
(Helfand et al. 2001) inside a much larger nebula (Markwardt \& \"{O}gelman 
1995; Aharonian et al. 2006c). Also, a small extended X-ray feature was 
detected around PSR~J1826-1334 with {\it XMM-Newton}, as part of a more 
extended nebula (Gaensler et al. 2003), whose asymmetric profile is the key 
to establishing a physical connection between this PWN and HESS~J1825-137 
(Aharonian et al. 2005b). In this case, the X-ray emission is much less 
extended than the gamma-ray emission, which might be attributable to the
difference in the cooling times of the X-ray and gamma-ray emitting
electrons (Aharonian et al. 2005b). The same might also be true for 
CXOU~J180351.4-213707, if it is the X-ray counterpart of HESS~J1804-216. 
Moreover, PSR~J1826-1334 is offset from the gamma-ray peak of HESS~J1825-137, 
as would be the case for CXOU~J180351.4-213707/HESS~J1804-216 if the PWN 
scenario holds.

Assuming that CXOU~J180351.4-213707 is the X-ray counterpart of 
HESS~J1804-216, we proceeded to assemble the spectral energy distribution 
(SED) of the system. Figure~4 shows the results. Since
the X-ray emitting region could be much larger, as already discussed, 
the measured X-ray fluxes should only be taken as lower limits. 
We attempted to model the SED in a leptonic scenario, in which the X-ray 
emission is assumed to originate from the synchrotron radiation from 
relativistic electrons in the region and TeV gamma-ray emission from 
inverse Compton (IC) scattering of ambient photons by the same electrons.
The spectral distribution of the electrons is assumed to be of the form:
$Q_e \propto \gamma^{-s} e^{- \gamma / \gamma_{max}}$. We found a fairly
good fit to the X-ray SED with $s=0.4$ and $\gamma_{max}\simeq 10^8$,
for a spherical emitting region of radius $R=10^{17}$ $cm$ and magnetic 
field $B=3\mu G$, using the methodology described by Mastichiadis (1996).
The synchrotron self-Compton emission from such an electron spectrum peaks 
at frequencies well above $10^{28}~\rm Hz$. The position of the IC peak 
remains essentially unchanged with the addition of the cosmic microwave 
background (CMB) photons ($U_{CMB} = 0.33~\rm eV~cm^{-3}$), as shown
in Fig.~4, in conflict with the measured gamma-ray spectrum. Varying model
parameters does not fundamentally improve the situation. 

On the other hand, we could find a reasonable fit to the TeV gamma-ray 
spectrum (e.g., with $s \simeq 1.5$, $\gamma_{max} \simeq 10^7$, and 
$B \simeq 3 \mu G$), as also shown in Fig.~4. However, the required 
electron spectrum deviates significantly from that needed to explain the 
X-ray emission. This perhaps argues for a multi-zone scenario. 
Intriguingly, in this case, an extrapolation of the IC spectrum comes
very close to the X-ray measurements (see Fig.~4). Of course, one should
always keep in mind the possibility that, e.g., CXOU~J180351.4-213707 
might be the compact core of a much extended PWN and the true X-ray flux of 
the PWN might thus be much higher. More sophisticated modeling is beyond 
the scope of this work.

We conclude by noting the lack of X-ray emission from PSR~J1803-2137. It is 
a bit surprising that this Vela-like pulsar could not be detected in a 30 ks 
ACIS/{\em Chandra} observation. Given the measured $P$ ($=133$ ms) and 
$\dot{P}$ ($=1.34 \times 10^{-13}$ $s$ $s^{-1}$), the spin-down power of the 
pulsar ($\equiv -4\pi^2 I \frac{\dot{P}}{P^3}$, where $I$ is the moment of 
inertia of the neutron star) is 
about $2.25\times 10^{36}$ $erg$ $s^{-1}$. Becker \& Tr\"umper (1997) showed 
that rotation-powered pulsars typically radiate away 0.1\% of the spin-down 
power in the {\em ROSAT} band (0.1--2.4 keV). So, we would expect a 
0.1--2.4 keV luminosity of $2.25\times 10^{33}$ $erg$ $s^{-1}$ for 
PSR~J1803-2137, or a 
flux of $1.2\times 10^{-12}$ $erg$ $cm^{-2}$ $s^{-1}$, which is several orders 
of magnitude higher than the {\em Chandra} ($1\sigma$) upper limit of 
$1.7\times 10^{-15}$ $erg$ $cm^{-2}$ $s^{-1}$ for the total emission.

\begin{acknowledgements}
We wish to thank the anonymous referee for his/her very useful comments.
This research has made use of the NASA/IPAC Extragalactic Database (NED) 
which is operated by the Jet Propulsion Laboratory, California Institute 
of Technology, under contract with the National Aeronautics and Space 
Administration, and of the Simbad Database.
We gratefully acknowledge financial support from the Department of Energy.

\end{acknowledgements}

\begin{table}
\caption{Detected X-ray Sources$^\dag$}
\begin{tabular}{lllc}\hline\hline
Name & Right Ascension & Declination  & Count Rate \\
     & J2000           & J2000        & 10$^{-3}$ $cts$ $s^{-1}$ \\ \hline  
CXOU J180341.5-214034 & 18:03:41.54(2) & -21:40:34.8(2)   & 1.27 $\pm$ 0.22 \\
CXOU J180345.3-213038 & 18 03 45.34(1) & -21 30 38.7(2)   & 2.42 $\pm$ 0.37 \\
CXOU J180349.1-212317 & 18:03:49.19(3) & -21:23:17.2(4)   & 4.34 $\pm$ 0.79  \\
CXOU J180349.3-214135 & 18:03:49.35(2) & -21:41:35.8(2)   & 1.54 $\pm$ 0.26  \\
CXOU J180350.9-213837 & 18 03 50.94(1) & -21 38 37.8(1)   & 1.48 $\pm$ 0.25 \\ 
CXOU J180351.4-213707 & 18 03 51.411(4) & -21 37 07.37(5) & 4.25 $\pm$ 0.32 \\ 
CXOU J180355.0-213937 & 18 03 55.00(1) & -21 39 37.4(1)   & 1.12 $\pm$ 0.17 \\ 
CXOU J180400.7-214251 & 18 04 00.76(2) & -21 42 51.5(2)   & 2.41 $\pm$ 0.32 \\ 
CXOU J180401.2-213153 & 18 04 01.217(5) & -21 31 53.48(7) & 6.23 $\pm$ 0.53 \\ 
CXOU J180404.2-213709 & 18 04 04.27(1) & -21 37 09.5(1)   & 0.97 $\pm$ 0.19 \\ 
CXOU J180432.4-214009 & 18 04 32.47(2) & -21 40 09.8(3)   & 4.58 $\pm$ 0.45 \\ \hline
\end{tabular} \\
$^\dag$The coordinates and count rates shown are based on the output of 
{\it celldetect}, except for CXOU J180351.4-213707 (see text). The numbers 
in parentheses indicate uncertainty in the last digit. Note that only 
statistical uncertainties are shown. \\
\end{table}

\begin{figure}
\epsscale{1.}
\plotone{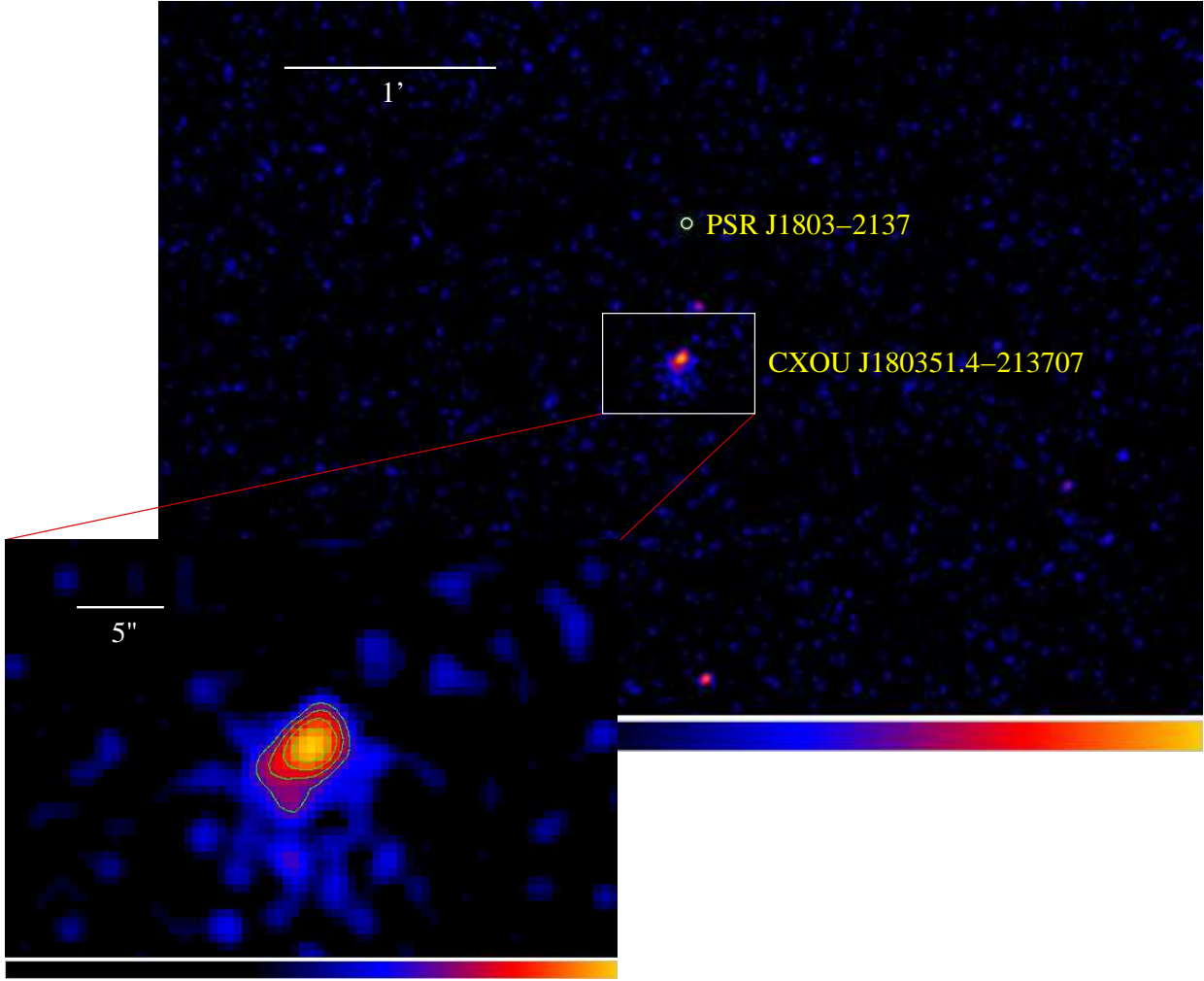}
\caption{X-ray View of PSR J1803-2137 and its surroundings. The image was made
in the 0.5--10 keV band. It has been smoothed and is shown on a logarithmic 
scale. The pulsar is not detected; its radio position is indicated by an 
open circle. On the other hand, the presence of CXOU J180351.4-213707 is 
apparent. The inset shows an expanded view of the source, with contours (at 
the levels of 0.50, 0.87, 1.52,and 2.64 counts) overlaid to show its 
extension. }
\end{figure}

\begin{figure}
\includegraphics[width=4.5in,angle=-90]{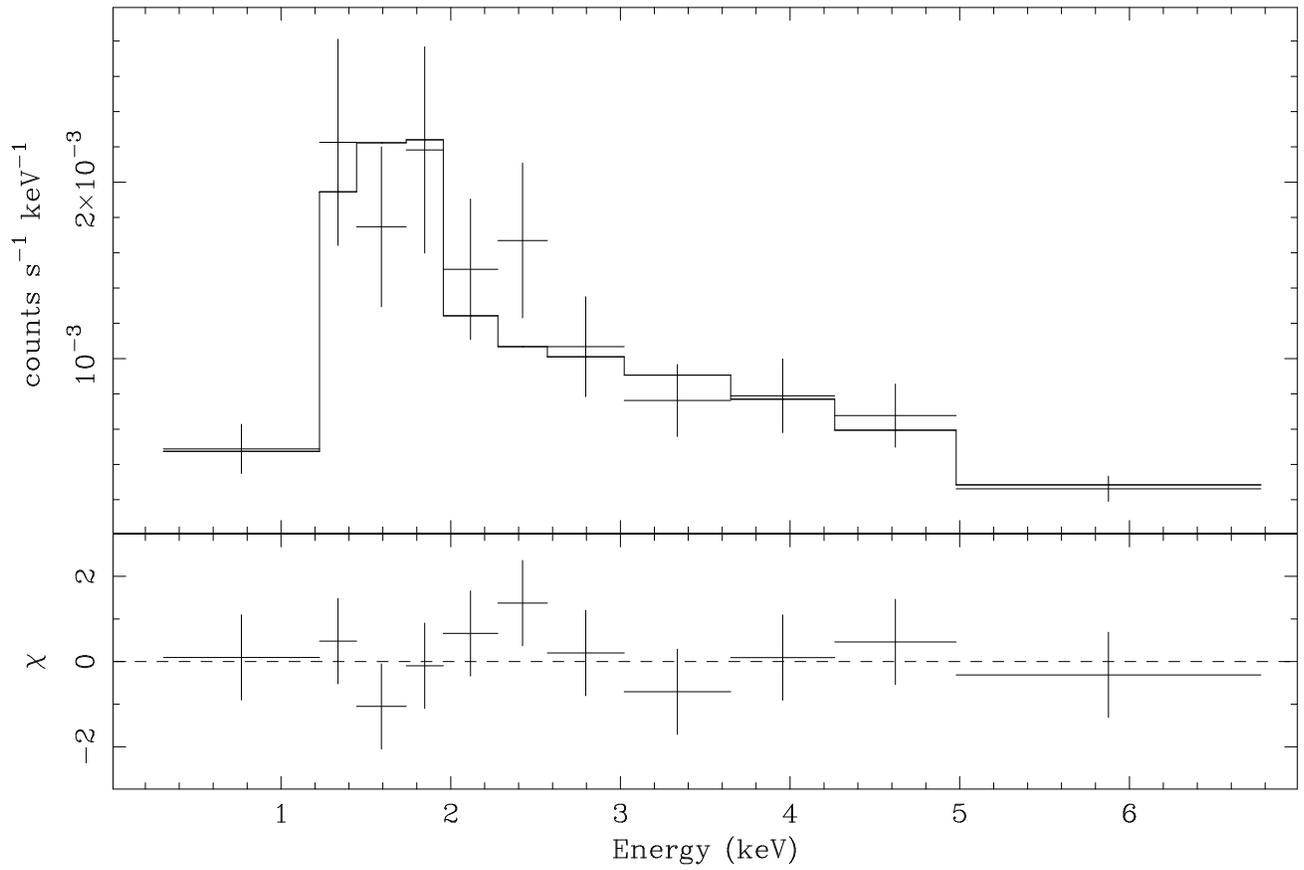}
\caption{X-ray spectrum of CXOU J180351.4-213707. The solid histogram in the
top panel shows the best fit to the data with an absorbed power-law model.
The residuals are shown in the bottom panel. }
\end{figure}

\begin{figure}
\epsscale{1.}
\includegraphics[width=4.5in]{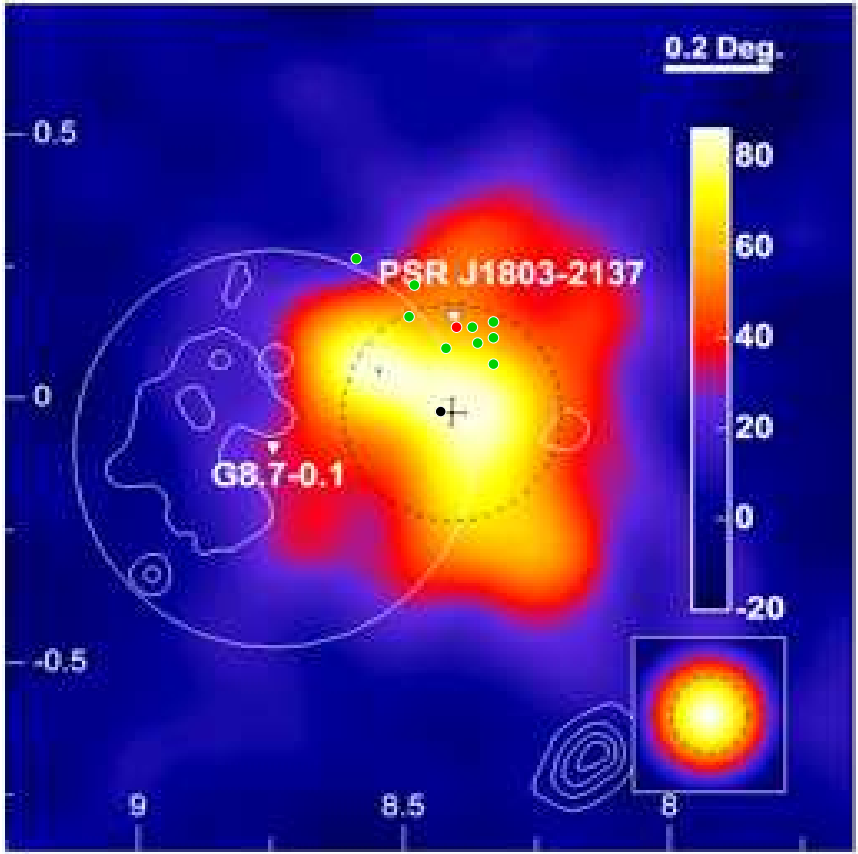}
\caption{X-ray sources in the vicinity of HESS J1804-216. The TeV gamma-ray
image was adapted from Aharonian et al. 2006a. The dashed circle is meant to
show the
average extent of the gamma-ray emission region. The positions of the 
detected X-ray sources are indicated by filled circles. Highlighted are 
CXOU J180351.4-213707 (in red), which might be the X-ray counterpart of 
HESS J1804-216, and CXOU J180432.4-214009 (in black), which is likely the 
same source as Suzaku J1804-2140. The {\it ROSAT} contours of G8.7-0.1
(Finley \& \"{O}gelman 1994) are also overlaid. The white circle shows the 
extent of the 20 cm emission as measured with the {\it VLA}. }

\end{figure}

\begin{figure}
\includegraphics[scale=0.75]{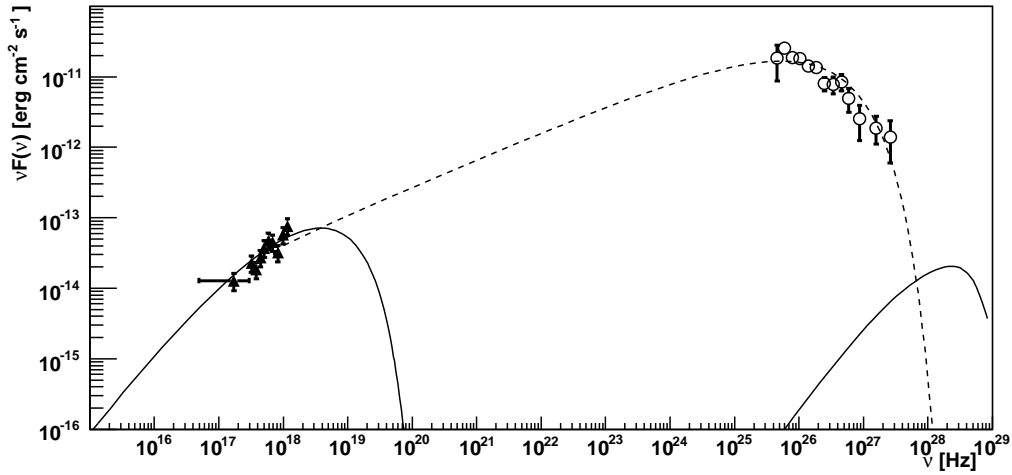}
\caption{The combined spectral energy distribution of CXOU~J180351.4-213707 
(in triangles) and HESS~J1804-216 (in open circles). The solid curve shows
representative results from leptonic calculations that aim at fitting 
the X-ray data, while the dashed curve shows those that fit the gamma-ray 
data (see text).
}
\end{figure}

\end{document}